\documentclass{jetpl}
\twocolumn

\lat
\usepackage[dvips]{graphicx}

\def\tlab{\mbox{$T_{\rm Lab}$}}
\def\tlabm{\mbox{$T_{\rm Lab}^{\rm max}$}}

\title{One\d rank interaction kernel of the two{\d}nucleon system
for medium and high energies}
\rtitle{One\d rank interaction kernel of the two{\d}nucleon system\ldots}
\sodtitle{One\d rank interaction kernel of the two{\d}nucleon system
for medium and high energies}

\author{S.\,G. Bondarenko, V.\,V. Burov, W{\d}Y. Pauchy Hwang$^{+}$,
and E.\,P. Rogochaya\/\thanks{e-mail: rogoch@theor.jinr.ru}}
\rauthor{S.\,G. Bondarenko, V.\,V. Burov, W{\d}Y. Pauchy Hwang,
and E.\,P. Rogochaya}
\sodauthor{Bondarenko, Burov, Pauchy Hwang,
and Rogochaya}

\address{Joint Institute for Nuclear Research, 141980
Dubna, Moscow region, Russia\\~\\
$^+$National Taipei University, Taipei 106, Taiwan}

\dates{\today}{*}

\abstract{A new version of the separable kernel
of the nucleon{\d}nucleon interaction in the Bethe{\d}Salpeter approach is presented.
The phase shifts are fitted to recent experimental data for singlet and uncoupled triplet
partial waves of the neutron{\d}proton scattering with total
angular momenta $J=0,1$.
The results are compared with other model calculations.}


\begin{document}

\maketitle

\section{INTRODUCTION}

The relativistic description of the interaction between two
nucleons can be carried out using the Bethe{\d}Salpeter (BS)
equation~\cite{BS}.
The BS formalism allows a covariant description of various reactions
including electromagnetic ones. The BS approach with the separable
kernel of interaction (as a relativistic analogue
of the Lippmann{\d}Schwinger equation with the separable
potential~\cite{plessas1,plessas2}) demonstrates a good description
of electromagnetic processes with the deuteron (such as the elastic
lepton{\d}deuteron scattering, deuteron electro{\d} and photodisintegration) and
light nuclei (deep inelastic scattering of leptons),
see the review~\cite{obzor} and references therein.
For instance, the BS formalism facilitates analysis of the role
of $P$ waves (negative energy partial{\d}wave components of the BS amplitude)
in the electromagnetic properties of the deuteron and its comparison
with the nonrelativistic treatment~\cite{hamamoto}.
Furthermore, the covariant BS approach
makes it possible to analyze off{\d}mass{\d}shell effects and contributions
of the relativistic two{\d}body currents.

The consistency of the consideration of the deuteron
breakup reactions demands the final state interaction (FSI)
between the outgoing nucleons to be taken into account.
To introduce the FSI of the final $np$ pair the BS equation for continuous
state should be solved. These calculations were performed
in the paper~\cite{manabe}
for the separable kernel of nucleon{\d}nucleon ($NN$) interaction with a relativistic generalization
of Yamaguchi functions. The obtained kernel allows one to describe the $NN$ phase
shifts up to laboratory kinetic energy \tlab $\sim 0.6{\ch}0.7$\,GeV
(for example, the nonrelativistic Bonn potential can
be applied up to \tlab $\sim 0.35$\,GeV). However, when the
amplitudes of reactions in a high energies region are calculated,
this type of functions leads to the presence of nonintegrable singularities.
Therefore there are many approaches called quasipotential where the two{\d}body
Green's function in the BS equation is replaced by a suitable function
which allows to reduce the 4{\d}dimensional integral to the 3{\d}dimensional
one, like in~\cite{sugar}.

We construct a relativistic generalization
of the modified Yamaguchi form factors. For scalar nucleons such functions
were introduced in~\cite{schwarz}.
Functions of this type do not contain poles on the real axis of the relative
energy. We use them for the description of the proton{\d}neutron scattering
in the wide region of \tlab (till 3\,GeV).
It is necessary when we consider
reactions with light nuclei with a high momentum transfer.
In particular, the calculations of the $d(e,e'p)n$ reaction are supposed
to be performed for the relativistic energies using the separable
functions elaborated in the paper (see, for example,
different kinematical conditions in the Table from~\cite{edenp}).

In this work the one{\d}rank kernel of the $NN$ interaction for singlet
and uncoupled triplet partial states of the neutron{\d}proton ($np$) scattering
with total angular momenta $J=0,1$ are presented. The paper is organized
as follows: in Sec.2 we set out basic formulae of the formalism, in
Sec.3 the functions under consideration are described,
in Sec.4 we show the scheme of the calculations and results,
Sec.5 is devoted to the analysis of the results and the comparison with calculations
within other models, and in Sec.6 we summarize the approach.

\section{BASIC FORMULAE}

Let us consider the partial{\d}wave decomposed BS equation for the
nucleon{\d}nucleon $T$ matrix in the rest frame of two particles
(the square of the total momentum
$s=(p_1+p_2)^2$ and the relative momentum $p=(p_1-p_2)/2$
[$p'=(p_1'-p_2')/2$] are defined through
the initial $p_1,~p_2$ and final $p_1',~p_2'$ nucleons 4{\d}momenta):
\begin{eqnarray}
&&T_{ll}(p_0',p';p_0,p;s)=V_{ll}(p_0',p';p_0,p;s)
\nonumber\\
&&\hskip 15mm+\frac{i}{4\pi^3} \int dk_0\int
k^2dk V_{ll}(p_0',p';k_0,k;s)\nonumber\\
&&\hskip 29mm\times S(k_0,k;s) T_{ll}(k_0,k;p_0,p;s).
\label{BS}
\end{eqnarray}
Here $T_{ll}$ is the partial{\d}wave decomposed $T$ matrix
($l$ enumerates $^SL_J^\rho$ states for simplicity, $S$ is the spin
of two particles, $L$ is an orbital momentum, and $\rho$ spin defines a sign of an energy),
$V_{ll}$ is the interaction kernel, $E_k=\sqrt{k^2+m^2}$, $m$ is a nucleon mass.
We consider uncoupled partial states, and therefore the $T$ matrix does not have
transitions to different partial states.
The two{\d}particle propagator is
\begin{eqnarray}
S(k_0,k;s) = 1/\left((\sqrt s/2-E_k+i\epsilon)^2-k_0^2\right).
\label{prop}
\end{eqnarray}

Using the separable anzatz for the one{\d}rank kernel of the $NN$
interaction
\begin{eqnarray}
V_{ll}(p_0',p';p_0,p;s)=\lambda_l(s) g^{[l]}(p_0',p')g^{[l]}(p_0,p),
\end{eqnarray}
we obtain the solution of the equation (\ref{BS}) for the $T$ matrix in the
similar separable form:
\begin{eqnarray}
T_{ll}(p_0',p';p_0,p;s)=\tau_l(s)g^{[l]}(p_0',p')g^{[l]}(p_0,p),
\end{eqnarray}
where
\begin{eqnarray}
\tau_l(s)=1/(\lambda_l(s)^{-1}+h_l(s))
\end{eqnarray}
with
\begin{eqnarray}
h_l(s)=-\frac{i}{4\pi^3}\int dk_0\int
k^2dk \frac{(g^{[l]}(k_0,k))^2}{(\sqrt
s/2-E_k+i\varepsilon)^2-k_0^2},
\end{eqnarray}
and $\lambda_l$ is a model parametric function. From the
normalization condition for the $T$ matrix in the on{\d}mass{\d}shell form:
\begin{eqnarray}
T_{ll}(s)\equiv T_{ll}(0,\bar p;0, \bar p;s)=-\frac{16\pi}{\sqrt
s\sqrt{s-4m^2}}e^{i\delta_l}\sin\delta_l, \label{T_norm}
\end{eqnarray}
where $\bar p=\sqrt{s/4-m^2}=\sqrt{m \tlab/2}$ is the
on{\d}mass{\d}shell momentum, the phase shift $\delta_l$ for the one{\d}rank
kernel of interaction is defined as
\begin{eqnarray}
\cot\delta_l(s)=\frac{\Real{T_{ll}(s)}}{\Imag{T_{ll}(s)}}=
-\frac{\lambda_l(s)^{-1}+\Real{h_l(s)}}{\Imag{h_l(s)}}.
\label{phases}
\end{eqnarray}

We obtain the low{\d}energy scattering parameters expanding
the following expression for $S$ waves into series of $\bar p${\d}terms \cite{low-en-exp}:
\begin{eqnarray}
\bar p\cot\delta_l(s)=-1/a_l+\frac{1}{2}\bar p^2 r_{0l}
+{\cal O}(\bar p^3).
\label{low}
\end{eqnarray}
Here the scattering length $a_l$ and the effective range $r_{0l}$ are introduced.
Since it is clear which partial states are discussed, hereinafter
we omit the index $l$ for simplicity.

\section{ONE{\d}RANK KERNEL}

We analyze two relativistic generalizations
of the Yamaguchi form factors: modified Yamaguchi (MY) functions
and modified extended  Yamaguchi (MEY) functions.

\subsection{Modified Yamaguchi functions}

For the description of the chosen partial states we use the
following covariant expressions:
\begin{eqnarray}
g^{[S]}(p_0,p)=\frac{(p_{c1}-p_0^2+p^2)}{(p_0^2-p^2-\beta^2)^2+\alpha^4},
\label{simple_s}
\end{eqnarray}
\begin{eqnarray}
g^{[P]}(p_0,p)=\frac{\sqrt{-p_0^2+p^2}}{(p_0^2-p^2-\beta^2)^2+\alpha^4}.
\label{simple_p}
\end{eqnarray}
The functions are numerated by angular momenta $L=0([S]),1([P])$.
The numerator in $g^{[S]}$ is introduced to compensate an
additional dimension in the denominator to provide the total
dimension as GeV$^{-2}$. This form was chosen because at
$p_{c1}=\beta^2$, $\alpha=0$ we get the function $g^{[S]}$ in the
standard Yamaguchi form~\cite{manabe}. We prefer not to
consider the case with the square root in the denominator as it
is used in~\cite{schwarz} for $S$ waves to avoid the calculations
with cuts on the real axis in $p_0$ complex plane.

\subsection{Modified extended Yamaguchi functions}

To extend the form of functions with increasing number of parameters
we introduce the following $g$ functions:
\begin{eqnarray}
g^{[S]}(p_0,p)=&&\frac{(p_{c1}-p_0^2+p^2)}{(p_0^2-p^2-\beta_1^2)^2+\alpha_1^4}
\nonumber\\
&&+\frac{C_{12}(p_0^2-p^2)(p_{c2}-p_0^2+p^2)^2}
{((p_0^2-p^2-\beta_2^2)^2+\alpha_2^4)^2},
\label{ext_s}
\end{eqnarray}
\begin{eqnarray}
g^{[P]}(p_0,p)=&&\frac{\sqrt{-p_0^2+p^2}}{(p_0^2-p^2-\beta_1^2)^2+\alpha_1^4}
\nonumber\\
&&+\frac{C_{12}\sqrt{(-p_0^2+p^2)^3}(p_{c3}-p_0^2+p^2)}
{((p_0^2-p^2-\beta_2^2)^2+\alpha_2^4)^2}.
\label{ext_p}
\end{eqnarray}
The denominators with $p_{c1}, p_{c2}, p_{c3}$ are intended for dimension
compensation as $p_{c1}$ for $g^{[S]}$ in the previous case~(\ref{simple_s}).

\section{CALCULATIONS AND RESULTS}

Using the $np$ scattering data we analyze the parameters of the separable kernel
distinguishing three different cases:
\begin{enumerate}
\item
There are no sign change in phase shifts or bound state ($^1P_1,~^3P_1$ partial states).
In this case
\begin{eqnarray}
\lambda_l(s)=const.
\end{eqnarray}
This is sufficient for the most of the higher partial waves.
\item
One sign change and no bound state ($^1S_0,~^3P_0$ partial states).
In this case the energy{\d}dependent expression for $\lambda_l$ is used
(see \cite{schwarz} and references therein):
\begin{eqnarray}
\lambda_l(s) = \lambda (s_0-s).
\end{eqnarray}
Here the parameter $s_0$ is introduced to reproduce the sign change in
the phase shifts at the position of the experimental value for the kinetic energy
\tlab\ where they are equal to zero. It is added to the other
parameters of the kernel.
\item
One sign change and one bound state ($^3S_1$ state).
In this case in addition to the zero in the phase shifts the $T$ matrix
has a pole at the mass of the bound state $M_b$:
\begin{eqnarray}
\det|\tau^{-1}(s=M_b^2)|=0.
\label{tauz}
\end{eqnarray}
To achieve this we set
\begin{eqnarray}
\lambda_l(s) = \lambda\frac{s_0-s}{s-m_0^2}. \label{l3}
\end{eqnarray}
The parameter $m_0$ is chosen to satisfy the condition
(see eq.~(\ref{tauz}) and~(\ref{l3}))
\begin{eqnarray}
\lambda^{-1}\frac{M_b^2-m_0^2}{s_0-M_b^2}+\Real{h(M_b^2)}=0,
\end{eqnarray}
\end{enumerate}
where $M_b=(2m-E_b)$ defines the mass of the bound state and $E_b$ is its energy.

The calculation of the parameters is performed by using
the equations~(\ref{phases}), (\ref{low}) and expressions
given in previous section
to reproduce experimental values for the phase shifts
till the maximal laboratory kinetic energy \tlab $<$ \tlabm,
deuteron energy and low{\d}energy scattering parameters.
Experimental data for phase shifts are taken from the
SAID program (http://gwdac.phys.gwu.edu) and for the rest quantities {\t} from~\cite{low-en}.
For all states except $~^3P_0$
the \tlabm\, is chosen to be $\sim 1$\,GeV both for MY and MEY cases
and for $~^3P_0$ wave {\t} $\sim 3$\,GeV for MEY case.

Now we find the introduced parameters of the kernel:
\begin{enumerate}
\item
For $P$ waves the minimization procedure for the function
\begin{eqnarray}
\chi^2=\sum\limits_{i=1}^{n} (\delta^{\rm
exp}(s_i)-\delta(s_i))^2/(\Delta\delta^{\rm exp}(s_i))^2
\label{mini_p}
\end{eqnarray}
is used. Here the number of experimental points $n$
depends on the \tlabm\, value.
\item
For $S$ waves the values of the scattering length
$a$ are also included into the minimization procedure
\begin{eqnarray}
\chi^2=&&\sum\limits_{i=1}^{n} (\delta^{\rm
exp}(s_i)-\delta(s_i))^2/(\Delta\delta^{\rm exp}(s_i))^2
\nonumber\\
&&+(a^{\rm exp}-a)^2/(\Delta a^{\rm exp})^2.
\label{mini_s}
\end{eqnarray}
The effective range $r_0$ is calculated through the obtained
parameters and compared with the experimental value $r_0^{\rm exp}$.
The value of the binding energy $E_d$ for the case $^3S_1$ is
taken into account by introduction of the parameter $m_0$ in
eq.(\ref{l3}).
\end{enumerate}

The calculated parameters of the one{\d}rank kernel
with MY and MEY functions and \tlabm\,
are listed in Tables~1
(here the values of $s_0$
are presented, too) and 2.
In Table 3 the calculated
low{\d}energy scattering parameters for $S$ waves are compared with their
experimental values.

In Figs.\ref{1p1}{\ch}\ref{3s1} the results of the phase
shifts calculations are compared with experimental data and
two alternative descriptions by CD{\d}Bonn \cite{bonn} and SP07
\cite{SP07}.

\section{DISCUSSION}

\begin{figure}
\begin{center}
\includegraphics[width=135mm]{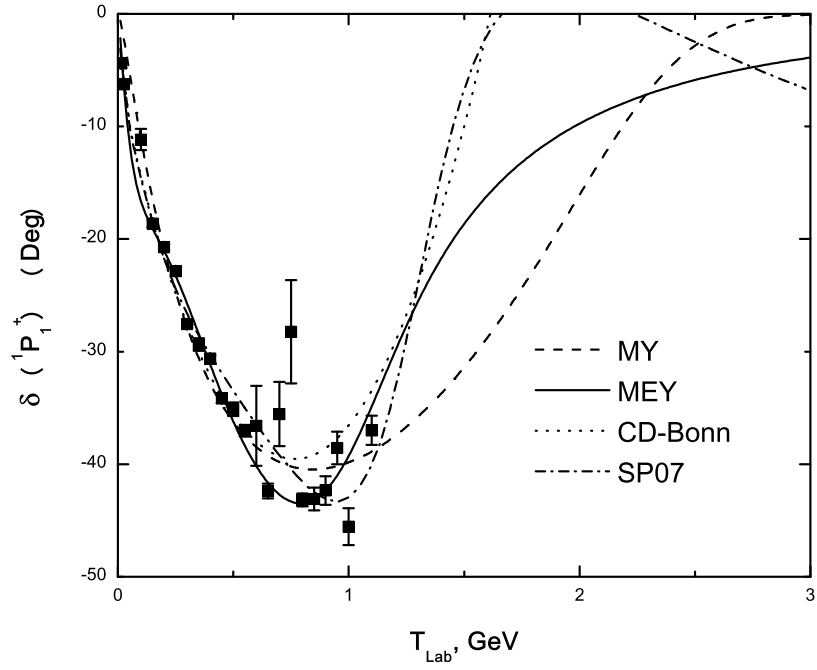}
\caption{{Phase shifts for the $^1P_1^+$ wave. Dashed line corresponds
to our parametrization with MY functions, eq.(\ref{simple_p}); the solid line
illustrates the extended case, eq.(\ref{ext_p}); the dotted and
dash{\d}dotted lines describe the CD{\d}Bonn and SP07 calculations,
correspondingly.}}
\label{1p1}
\end{center}
\end{figure}

The calculated phase shifts are presented in Figs.\ref{1p1}{\ch}\ref{3s1} and
parameters of the separable kernel are given in Tables~1{\ch}2
(MY case is denoted by the dashed line, MEY one {\t} by the solid line).
The calculations with the nonrelativistic CD{\d}Bonn potential (dotted line) and the solution SP07
\cite{SP07} (dash{\d}dotted line) are included for comparison.

We can see in Fig.\ref{1p1} that all the calculations give
a reasonable agreement with experimental data up to energy
\tlab $\sim 1.1$\,GeV but at larger energies their behavior becomes different
drastically. Note that the CD{\d}Bonn and SP07 results for phase shifts even change
sign at $1.5< \tlab < 2.5$\,GeV. So we can stress that it is very desirable to obtain
more exact determination of experimental data for the $^1P_1^+$ channel in this energy range.
Now keeping in mind that the discrepancies between CD{\d}Bonn, MEY and
SP07 calculations at $\tlab< 1.2$\,GeV are not so large we can trust our
calculations in this range of energies.

\begin{figure}
\begin{center}
\includegraphics[width=135mm]{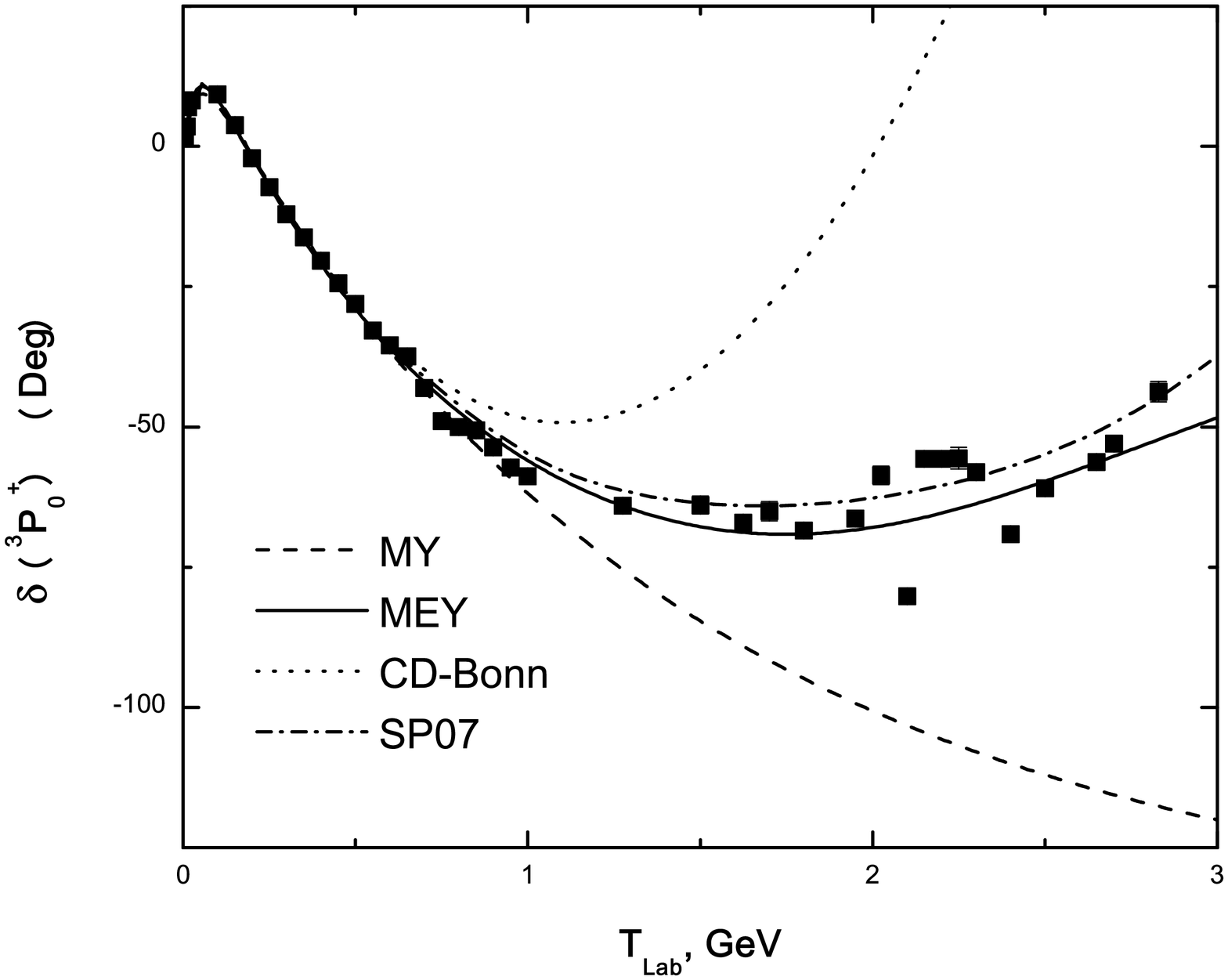}
\caption{{The same as in Fig.1 for the $^3P_0^+$ state.}}
\label{3p0}
\end{center}
\end{figure}

In Fig.\ref{3p0} we can see
that CD{\d}Bonn and MY calculations demonstrate an opposite behavior for $^3P_0^+$
phase shifts at $\tlab >1$\,GeV. The MEY and SP07 calculations
give a very good agreement with experimental data in a wide range of energies
$\tlab < 3$\,GeV. Thus, our model for the kernel of the $NN$ interaction
in the $^3P_0^+$ channel is acceptable to be used in various relativistic calculations
of reactions.

\begin{figure}
\begin{center}
\includegraphics[width=135mm]{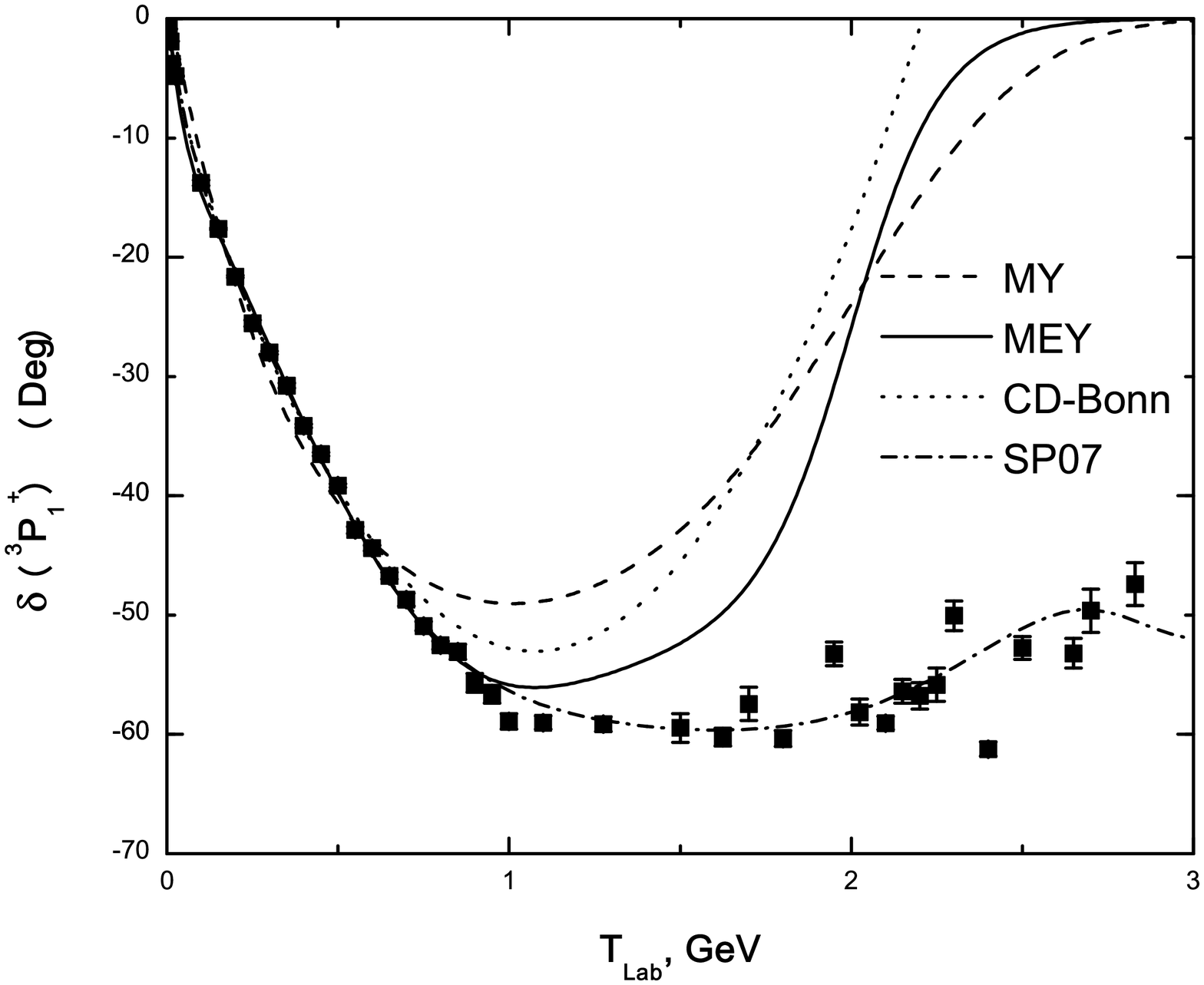}
\caption{{The same as in Fig.\ref{1p1} for the $^3P_1^+$ state.}}
\label{3p1}
\end{center}
\end{figure}

Calculations for the $^3P_1^+$ channel
(see Fig.\ref{3p1}) show us that MY, CD{\d}Bonn, and MEY have similar
behavior in a wide energy range, but can explain experimental data up
to $\tlab=$0.55, 0.6 and 1.2\,GeV, respectively.
The SP07 calculations give a suitable agreement with experimental data up
to 3\,GeV. Thus to explain the data for the $^3P_1^+$ channel
in wider energy range
in our approach we need to increase the rank of the separable kernel
of the $NN$ interaction.

\begin{figure}
\begin{center}
\includegraphics[width=135mm]{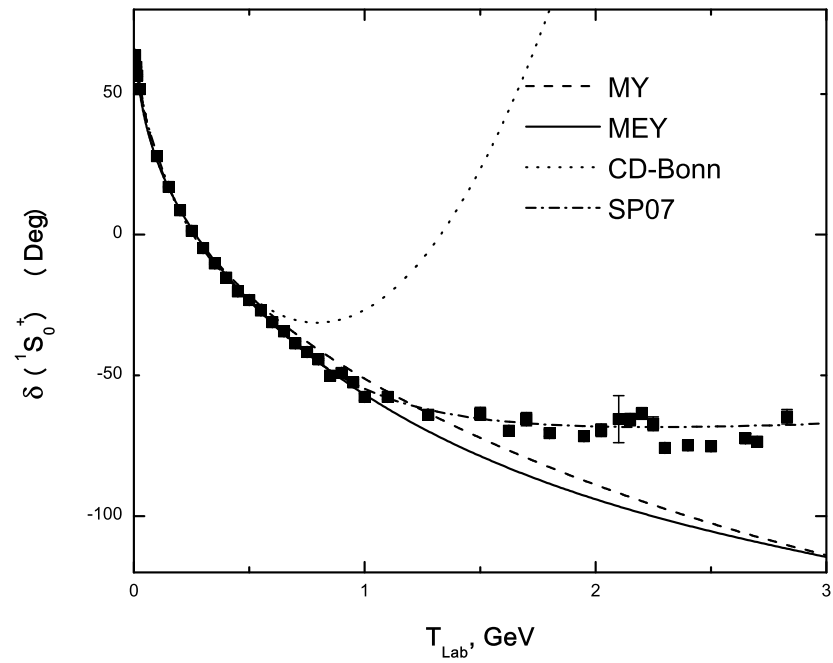}
\caption{{The same as in Fig.\ref{1p1} for the singlet partial state
$^1S_0^+$, but the description by MY functions is defined by eq.(\ref{simple_s})
and extended one (MEY) {\t} by eq.(\ref{ext_s}).}} \label{1s0}
\end{center}
\end{figure}

\begin{figure}
\begin{center}
\includegraphics[width=135mm]{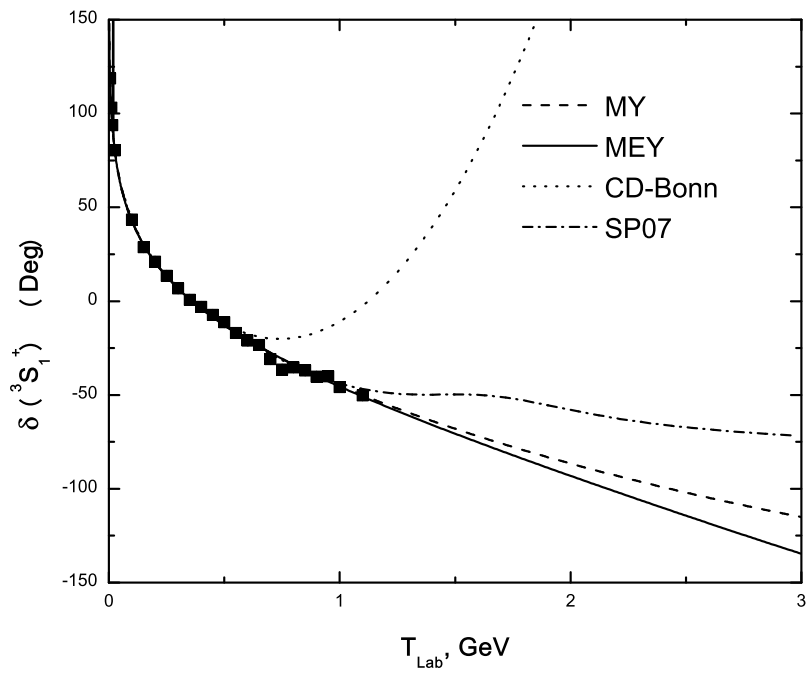}
\caption{{The same as in Fig.\ref{1s0} for the triplet partial state
$^3S_1^+$.}} \label{3s1}
\end{center}
\end{figure}

Fig.\ref{1s0} demonstrates a close agreement with experimental data for CD{\d}Bonn up to the energy
$\tlab < 0.6$\,GeV, MY and MEY up to $\tlab < 1.2$\,GeV in the $^1S_0^+$
channel
(note that these calculations give very similar behavior of phase shifts
in the wide energy range up to 3\,GeV). We can conclude that in this case
as well as for $^3P_1^+$ we need to increase the rank of the separable kernel
of the $NN$ interaction. The SP07 calculations describe experimental data for the $^1S_0^+$ channel up to 3\,GeV.

Fig.\ref{3s1} for the $^3S_1^+$ wave shows us that CD{\d}Bonn calculations agree
with experimental data up to 0.6\,GeV. MY, MEY, and SP07 have
similar behavior in a wide energy range and explain known experimental data
up to $\tlab =$ 1.1\,GeV. It should be noted that calculations of
low{\d}energy parameters with MY and MEY form factors
give us a reasonable agreement with their
experimental values (see Table 1).
%\ref{spar}

\section{CONCLUSION}
We present the new parametrizations of the separable kernel
of the $NN$ interaction which are adopted for calculations at large energies.
As it was expected the MEY functions give a better description
of the scattering data than the MY ones. Using the suggested MEY form factors
the phase shifts are described in a whole range
of measured energies for the following partial states: $^1P_1^+$, $^3S_1^+$
($\tlab <$ 1.2\,GeV) and  $^3P_0^+$ ($\tlab <$ 3\,GeV).
The phase shifts for the $^1S_0^+$ and $^3P_1^+$ partial states can be described
in our approach up to $\tlab <$ 1.2\,GeV. To improve an agreement
with experimental data up to $\tlab <$ 3\,GeV it is necessary to increase
the rank of the separable kernel of the $NN$ interaction.
It is planned to be done in the near future.

\section*{ACKNOWLEDGEMENTS}

We wish to thank our collaborators A.\,A.~Goy,
K.\,Yu.~Kazakov, D.\,V.~Shulga and Y.~Yanev
for fruitful discussions.

\newpage
\onecolumn

\begin{center}
\begin{table}
\caption{Parameters of the one{\d}rank kernel with modified Yamaguchi functions.}
\centering
\begin{tabular}{lcc|lccc}
\hline\hline
           & $^1S_0^+$ & $^3S_1^+$ &                     & $^1P_1^+$ & $^3P_0^+$ & $^3P_1^+$ \\
\tlabm     & 0.999\,GeV & 1.1\,GeV   &                     & 1.1\,GeV   & 0.999\,GeV & 0.999\,GeV \\
\hline
$\lambda$ (GeV$^2$) & -1.0375   &  -10.004  & $\lambda$ (GeV$^4$) & 8.11284   & -211.74   &  11.962  \\
$\alpha$  (GeV)     &  1.3312   &  -11.287  & $\beta$   (GeV)     & 0.56487   & 0.53808   &  0.58466  \\
$\beta$   (GeV)     &  0.1      &   1.3540  & $\alpha$  (GeV)     & 0.2       & 0.85720   &  0.2      \\
$p_{c1}$  (GeV$^2$) &  19.229   &   0.49696 &&&&\\
\hline
$s_0$     (GeV$^2$) &  4.0279   &   4.2020  &                     &           & 3.8682    &           \\
\hline\hline
\end{tabular}\label{spar}
\end{table}
\end{center}

\begin{center}
\begin{table}
\caption{Parameters of the one{\d}rank kernel with modified extended
Yamaguchi functions.}
\centering
\begin{tabular}{lcc|lccc}
\hline\hline
              & $^1S_0^+$ & $^3S_1^+$ &                      & $^1P_1^+$ & $^3P_0^+$ & $^3P_1^+$     \\
\tlabm        & 0.999\,GeV & 1.1\,GeV   &                      & 1.1\,GeV   & 2.83\,GeV  & 0.999\,GeV     \\
\hline
$\lambda$    (GeV$^2$) & -0.84694  & -9.1434   & $\lambda$    (GeV$^4$) & 0.054821  & -192.60   &   0.030177 \\
$C_{12}$     (GeV$^0$) & -8.6404   &  3.3641   & $C_{12}$     (GeV$^0$) & 4.67839   &  4.9604   &  -0.39145  \\
$\beta_{1}$  (GeV)     &  0.13400  &  2.6609   & $\beta_{1}$  (GeV)     & 0.21829   &  0.71329  &   0.19951  \\
$\beta_{2}$  (GeV)     &  0.92330  & 0.12501   & $\beta_{2}$  (GeV)     & 0.20829   &  0.70329  &   0.66216  \\
$\alpha_{1}$ (GeV)     &  1.4553   &  2.0808   & $\alpha_{1}$ (GeV)     & 0.20500   &  0.70958  &   0.20000  \\
$\alpha_{2}$ (GeV)     &  0.93129  &  1.6340   & $\alpha_{2}$ (GeV)     & 0.52835   &  1.4574   &   0.32015  \\
$p_{c1}$     (GeV$^2$) &  24.261   &  189.85   & $p_{c3}$     (GeV$^2$) & 0.36868   & -1.7576   &  -117.75   \\
$p_{c2}$     (GeV$^2$) & -2.1076   &  3.6688   & &&&\\
\hline\hline
\end{tabular}\label{ext_param}
\end{table}
\end{center}

\begin{center}
\begin{table}
\caption{The low{\d}energy scattering parameters and binding energy for the
singlet $(s)$ and triplet $(t)$ $S$ waves.} \centering
\begin{tabular}{lcccccccc}
\hline\hline
           &$a_{s}$(Fm) & $r_{0s}$(Fm) &$a_{t}$(Fm) & $r_{0t}$(Fm) &$E_d$(MeV)     \\
\hline
MY         &-23.80      & 2.4          & 5.43       & 1.8          & 2.2246       \\
MEY        &-23.74      & 2.51         & 5.39       & 1.73         & 2.2246       \\
Experiment &-23.748(10) & 2.75(5)      & 5.424(4)   & 1.759(5)     & 2.224644(46) \\
\hline\hline
\end{tabular}\label{lep}
\end{table}
\end{center}

\end{document}